\documentclass{SCIYA2017enOL}
\online
\begin{document}

\ensubject{fdsfd}

\ArticleType{ARTICLES}
\Year{2017}
\Month{January}%
\Vol{60}
\No{1}
\BeginPage{1} %
\DOI{10.1007/s11425-000-0000-0}
\ReceiveDate{January 1, 2017}
\AcceptDate{January 1, 2017}

\title[]{Data-based prediction and causality inference of nonlinear dynamics}
{Data-based prediction and causality inference of nonlinear dynamics}

\author[1]{MA Huanfei}{hfma@suda.edu.cn}
\author[2]{LENG Siyang}{syleng@sat.t.u-tokyo.ac.jp}
\author[2,3,$\ast$]{CHEN Luonan}{lnchen@sibs.ac.cn}



\address[1]{School of Mathematical Sciences, Soochow University, Suzhou {\rm 215006}, China}
\address[2]{Institute of Industrial Science, The University of Tokyo, Tokyo 153-8505, Japan}
\address[3]{Key Laboratory of Systems Biology, Innovation Center for Cell Signaling Network, \protect\\ Institute of Biochemistry and Cell Biology, Shanghai Institutes for Biological Sciences, \protect\\ Chinese Academy of Sciences,
Shanghai {\rm 200031}, China}

\abstract{Natural systems are typically nonlinear and complex, and it is of great interest to be able to reconstruct a system in order to understand its mechanism, which can not only recover nonlinear behaviors but also predict future dynamics. Due to the advances of modern technology, big data becomes increasingly accessible and consequently the problem of reconstructing systems from measured data or time series plays a central role in many scientific disciplines. In recent decades, nonlinear methods rooted in state space reconstruction have been developed, and they do not assume any model equations but can recover the dynamics purely from the measured time series data. In this review, the development of state space reconstruction techniques will be introduced and the recent advances in systems prediction and causality inference using state space reconstruction will be presented. Particularly, the cutting-edge method to deal with short-term time series data will be focused on. Finally, the advantages as well as the remaining problems in this field are discussed.}



\maketitle

\section{Introduction}
In the big data era, various types of time-series data are increasingly accumulated every day in various disciplines and in various scopes  \cite{lockhart2000genomics,de2002modeling,stein2013ecological,rienecker2011merra,fan2014challenges}. An outstanding problem is how to make system-level study based on the vast amount of data. Particularly, based on the measured/observed time series data,  how to reconstruct the system and further study the mechanism behind the phenomenon has become an attracting problem and a hot topic in interdisciplinary science. Despite of the widely used approaches employing linear models \cite{hamilton1994time}, which are mainly based on correlation and regression, the natural systems usually show nonlinear nature: the components within one system intertwine with each other, and the dynamics of each component cannot be separated from one another. Such characteristic property distinguishes nonlinear systems from the linear stochastic systems, and nonlinear methods are thus more favorable in reconstructing complex systems.

Since the embedding theory proposed by Takens in 1980s \cite{Takens1981}, the State Space Reconstruction (SSR) technique has been developed and applied in nonlinear time-series analysis. The key idea of the SSR stems from the non-separability of the nonlinear system. The property of non-separability implies that the dynamics of one observable variable could contain all the dynamical information of the whole system. As a model-free method, the SSR based methods have been attracting persistent interest and under rapid development in the recent decades \cite{hegger1999practical,Kantz2004}. In this review, we will not cover the whole topics of system reconstructions using SSR methods, which can be referred to some excellent reviews \cite{hegger1999practical,Kantz2004,small2005applied,causalreview2007,wang2016data}, but focus on the forefront of the recent advances in predicting future dynamics and inferring causality using SSR framework. Particularly, the majority of SSR based methods usually require sufficiently long time-series data or a large number of samples, while the real data sets always have limited length due to the experimental restrictions or costs. We will introduce in detail the recently developed methods aiming to deal with short-term time series data.

This review is organized in the following manner. In Section \ref{sec.SSR}, we first introduce the principle of various embedding theorems, which constitute the theoretical foundation of SSR. Based on the embedding theory, the framework of SSR as well as several related aspects in real applications are presented. In Section \ref{sec.prediction}, the developments of predicting nonlinear system using SSR in the recent decades are reviewed. Particularly, how to make predictions from short-term high-dimensional time series data will be focused and introduced in details. Section \ref{sec.Causality} presents the directional interaction detection methods, i.e., the causality inference methods using SSR framework. Both the recently proposed Convergent Cross Mapping (CCM) method, using convergent of mutual predictions,  and Cross Map Smoothness (CMS) method, particularly designed for short-term data, are introduced in details. We also speculate on several open problems, for instance, conditional causality and causal relations in delayed systems in this section. Conclusions and discussions can be found in Section \ref{sec.Conclusion}.

\section{State Space Reconstruction}
\label{sec.SSR}
\subsection{Embedding theory}
Consider a general dynamical system in the form of
\begin{equation}\label{dynamical system}
  \frac{{\rm d}\bm{x}}{{\rm d}t}=F(\bm{x}),
\end{equation}
where $\bm{x}=[x_1,x_2,\dots,x_n]^{\rm T}$ is the system's state vector and $F=[f_1,f_2,\dots,f_n]^{
\rm T}$ with each $f_i(\cdot)$ a nonlinear function.
Starting from the initial value $\bm{x}(0)=\bm{x}_0$, the trajectory $\bm{x}(t)$ describes the evolution of the system along time. After transient dynamics, the system's bounded trajectory will converge to an attractor, i.e., a bounded and invariant set $\mathcal{A}$. Here the diverging situations are not under consideration.

For a discrete system, the dynamical system is in the form of
\begin{equation}\label{dynamical system discrete}
\bm{x}({t+1})=F(\bm{x}(t)),
\end{equation}
where $F$ is a nonlinear map, and $\bm{x}(t), t=0,1,2,\dots$ is the trajectory of the system, and the attractor can be defined analogously.

We consider the dynamics constrained in a compact manifold $\mathcal{M}$ containing the attractor $\mathcal{A}$. For both continuous and discrete cases, the time evolution corresponding with an initial value $\bm{x}_0\in\mathcal{M}$ can be denoted uniformly by $\bm{x}(t)=F_t(\bm{x}_0)$: in the discrete case time $t\in\mathbb{N}$ and $F_t=(F)^t$, while in the continuous case time $t\in\mathbb{R}$ and $F_t(\bm{x}_0)$ is the integral curve through $\bm{x}_0$.

\begin{figure}
\begin{center}
\includegraphics[width=0.9\textwidth]{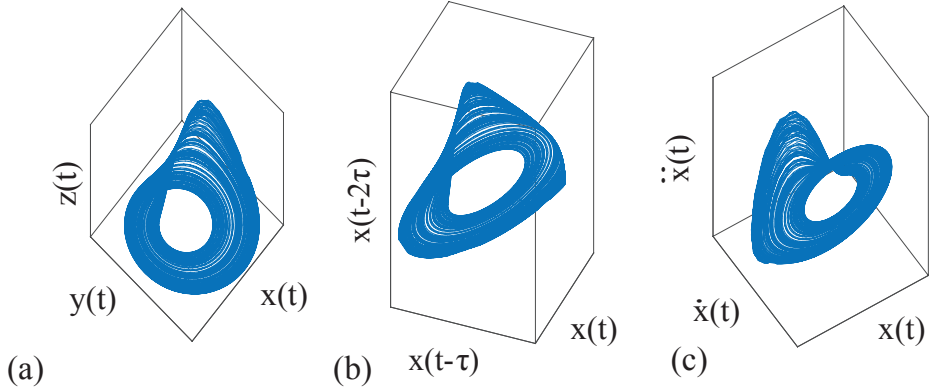}
\caption{(a) The attractor of system (\ref{rossler_packard}) in the original state space. (b)(c) Reconstructed attractor by single  time series.} \label{f1}
\end{center}
\end{figure}

Intuitively, to study the property of the attractor in the state space, the values of all the system's state components $x_i(t), i=1,2,\dots,n$ along time should be known.
However, the embedding theory implies that from a few or even one single time series observed from the system, it is possible to study the dynamical property of the whole system, i.e., the system can be reconstructed by observations.
Actually, it was first discovered by Packard \textit{et al} that using only one single coordinate of R\"{o}ssler system one can reconstruct a faithful state-space representation of the dynamics in the original 3-dimensional space \cite{Packard1980PRL}.

To be specific, the dynamical system of interest in \cite{Packard1980PRL} can be expressed in the form of Eq.(\ref{dynamical system}) as
\begin{equation}\label{rossler_packard}
  \begin{array}{l}
    \displaystyle\frac{{\rm d}x}{{\rm d}t}=-(y+z),\\
    \displaystyle\frac{{\rm d}y}{{\rm d}t}=x+0.2y,\\
    \displaystyle\frac{{\rm d}z}{{\rm d}t}=0.4+xz-5.7z,\\
  \end{array}
\end{equation}
and with time series of all the state components, the system's attractor can be obtained, as shown in Fig.~\ref{f1}(a) under $x-y-z$ coordinates.
As suggested by Packard \textit{et al}, by sampling the time series of only single coordinate $x(t)$, a variety of three independent coordinate sets can be obtained such as $[x(t), x(t-\tau), x(t-2\tau)]$ and $[x(t), \dot{x}(t), \ddot{x}(t)]$ where $\tau$ is a sampling interval and $\dot{x},\ddot{x}$ are the first order and second order differences respectively. The attractors in such coordinates are shown in Fig.~\ref{f1}(b)(c) respectively. The geometry property of the attractors in Fig.~\ref{f1}(b)(c) are quite similar to that in Fig.~\ref{f1}(a). As a matter of fact, the dimensionality and the positive characteristic exponent can be further computed for each of these attractors \cite{Packard1980PRL} and the results confirm that the attractors obtained in Fig.~\ref{f1}(b)(c) can be regarded as reconstructions of the attractor obtained in Fig. \ref{f1}(a).

Such discovery was theoretically confirmed by Takens in his seminal work \cite{Takens1981}, in which the famous Takens embedding theory was proposed.
\begin{theorem}[Takens embedding theory I]
Let $\mathcal{M}$ be a compact manifold of dimension $d$. For pairs $(\phi,h),~\phi:\mathcal{M}\rightarrow\mathcal{M}$ a smooth diffeomorphism and $h:\mathcal{M}\rightarrow\mathbb{R}$ a smooth function, it is a generic property that the map $\psi_{\phi,h}:\mathcal{M}\rightarrow\mathbb{R}^{L}$, defined by $$\psi_{\phi,h}(\bm{x})=(h(\bm{x}),h(\phi(\bm{x})),\dots,h(\phi^{L-1}(\bm{x})))$$ is an embedding where the integer $L=2d+1$.\hfill$\blacksquare$
\end{theorem}

Here an embedding of $\mathcal{M}$ means a smooth diffeomorphism from $\mathcal{M}$ onto its image $\psi(\mathcal{M})$, i.e., a smooth one-to-one map which has a smooth inverse. The number $L=2d+1$ is the dimension of the space for the reconstructed attractor.

Actually, the embedding theory could be stated in other forms, such as derivative embedding and generalized embedding theorems \cite{Takens1981,whitney1936differentiable,Deyle2011Plosone} stated as follows.

\begin{theorem}[Takens embedding theory II]
Let $\mathcal{M}$ be a compact manifold of dimension $d$. For pairs $(\phi,h),~\phi:\mathcal{M}\rightarrow\mathcal{M}$ a smooth diffeomorphism and $h:\mathcal{M}\rightarrow\mathbb{R}$ a smooth function, it is a generic property that the map $\psi_{\phi,h}:\mathcal{M}\rightarrow\mathbb{R}^{L}$, defined by $$\psi_{\phi,h}(\bm{x})=(h(\bm{x}),\frac{{\rm d}}{{\rm d}t}h(\phi(\bm{x})),\dots,\frac{{\rm d^{L-1}}}{{{\rm d} t}^{L-1}}h(\phi(\bm{x})))$$ is an embedding where the integer $L=2d+1$.\hfill$\blacksquare$
\end{theorem}

\begin{theorem}[Generalized embedding theory]\label{theorem3}
Let $\mathcal{M}$ be a compact manifold of dimension $d$ and a set of $L$ observation functions $\langle h_1,\dots,h_{L}\rangle$ where $h_k:\mathcal{M}\rightarrow\mathbb{R}$ a smooth function and $L=2d+1$. Then it is a generic property of all possible $\langle h_k\rangle$ that the map $\psi_{\phi,h}:\mathcal{M}\rightarrow\mathbb{R}^{L}$, defined by $$\psi_{\langle h_k\rangle}(\bm{x})=(h_1(\bm{x}),h_2(\bm{x})),\dots,h_L(\bm{x})))$$ is an embedding.\hfill$\blacksquare$
\end{theorem}

It is clear that here the dimension $d$ is important for the embedding theory. However, the dimension $d$ is usually much higher than the essential dimension of the attractor $\mathcal{A}$. Therefore, the embedding theory was later generalized into fractal dimensions by Sauer \textit{et al} in \cite{Sauer1991}.  Let $d_A$ be the box-counting dimension of the attractor $\mathcal{A}$, which could be a non-integer and could be much smaller than the integer dimension $d$ of the manifold $\mathcal{M}$ containing $\mathcal{A}$, then for almost every smooth measurement function $h$, the maps defined in the Takens embedding theorems are still an embedding on each compact subset of a smooth manifold contained within $\mathcal{A}$ with $L>2d_A$.

Particularly, starting from one dimensional observed time series, a delay coordinate map could be introduced.
\begin{definition}[Delay coordinate]
  If $\phi$ is a flow on a manifold $\mathcal{M}$, $\tau$ is a positive delay, and $h:\mathcal{M}\rightarrow \mathbb{R}$ is a smooth function, define the delay coordinate map $F(h,\phi,\tau):\mathcal{M}\rightarrow \mathbb{R}^L$ by
  $$F(h,\phi,\tau)(\bm{x})=(h(\bm{x}),h(\phi_{-\tau}(\bm{x})),h(\phi_{-2\tau}(\bm{x})),\dots,h(\phi_{-(L-1)\tau}(\bm{x})))$$\hfill$\blacksquare$
\end{definition}
According to the fractal embedding theorem, the delay coordinate map is generically an embedding as long as $L>2d_A$. Thus it is straightforward to see that when the observation function $h$ is simply chosen as coordinate function $x_i$, the delay coordinate map
$\bm{x}(t)\rightarrow[x_i(t),x_i(t-\tau),\dots,x_i(t-(L-1)\tau)]$ with sampling interval $\tau$ is generically an embedding.

The one-to-one property of embedding is essential to the reconstruction of the original system because the state of a deterministic dynamical system, as well as its future evolution, are completely specified by a point in the state space. Therefore, the delay coordinate map actually sets up a bridge between the dynamics of the original system and the dynamics in the reconstructed space. Thus the delay coordinate $[x_i(t),x_i(t-\tau),\dots,x_i(t-(L-1)\tau)]$ is a state space reconstruction of the original system, which implies that the one dimensional observed time series can contain all the information of the dynamics of the whole original system. Moreover, the embedding theory implies that the dynamics in the reconstructed space is topologically conjugated with the dynamics in the original state space \cite{Kantz2004}. Thus, the nonlinear dynamic properties such as Lyapunov exponent, entropy, and even the unstable periodic orbit (UPO) are preserved from the original system to the reconstructed system, which forms the cornerstone of the state space reconstruction technique.

\subsection{Issues related to state space reconstruction}
When applying delay coordinate SSR to real data sets, the information of the original nonlinear system is generally unknown \textit{a priori}, thus there are two key factors to be determined for successful SSR application: one is the embedding dimension $L$ and the other is the delay $\tau$. Dozens of methods for estimating optimal dimension $L$ and delay $\tau$ have been developed in the past few decades based on time series data \cite{Grassberger1983,Fraser1986PRA,Liebert1989PLA ,theiler1986spurious,liebert1991optimal,buzug1992optimal,FNN,kember1993correlation,rosenstein1994reconstruction,lai1996upper,lai1998effective,Pecora2007Chaos}.

Theoretically, any $\tau>0$ could be used in the delay coordinate SSR process, but in practice, the delay coordinates will be strongly correlated with too small $\tau$ while the reconstructed attractor can fold over on itself with too large $\tau$. Hence, one wants the smallest $\tau$ that maximizes the independence of the coordinates of the embedding vector. Therefore, several representative methods have been proposed to choose optimal $\tau$ by computing a statistic that measures the independence of $\tau-$separated points in the time series.  As $\tau$ increases, the first zero of the autocorrelation function of the time series, the first minima of the average mutual information, or that of the correlation sum \cite{Grassberger1983,Fraser1986PRA,Liebert1989PLA} could be candidates of optimal $\tau$. Actually, such $\tau$  maximizes general forms of independence, and consequently can be adopted in the delay coordinate.

Analogous to the case of choosing $\tau$, the embedding dimension $L$ should be neither too small nor too large. In theory, $L$ should be sufficiently large so as to satisfy the condition of $L>2d_A$, however in practice, low dimensional reconstruction is more favorable. On the other hand, it is noticed that $L>2d_A$ is a sufficient but not necessary condition. Thus the optimal $L$ should be chosen as the smallest value that affords a topologically correct result. To this, the false near neighbor (FNN) algorithm by Kennel \textit{et al} \cite{FNN} is a feasible method generally accepted. In an FNN-based method, one tries to construct a delay coordinate embedding with embedding dimension $k$, computes each point's near neighbors, and repeats such process with increased embedding dimension $k+1$. If any of the geometry relations changes, i.e., some neighbor in $k$ dimension is no longer a neighbor in $k+1$ dimension, then it is a false near neighbor and it indicates that the dynamics are not properly unfolded under embedding dimension $L=k$. Thus, the value appearing on the elbow of the decreasing FNN curve as $k$ increases can be a candidate for optimal embedding dimension.

In the recent years, it has been noted that the product $L*\tau$ is the factor that really matters. Thus the both factors can be estimated at the same time and a unified approach has been proposed \cite{Pecora2007Chaos}.

Furthermore, when applying SSR to practical measured time series, noise is unavoidable. The theory of state space reconstruction in the presence of noise was first considered in \cite{Casdagli1991}, which incorporated the observational noise and estimation error into the framework. Delay coordinate embedding theory for stochastic systems was later systematically studied by Stark \textit{et al} in \cite{Stark1997I} and \cite{Stark2003II}.

\section{Prediction of nonlinear dynamics}
\label{sec.prediction}
Predicting nonlinear dynamics using state space reconstruction method has a long history, which can be dated back to 1969 when Lorenz proposed the method of analogues in the state space to forecast atmosphere \cite{Lorenz1969predict}. Later the idea was shown to be applicable in the reconstructed state spaces by Pikovsky \cite{Pikovsky1986predict}. The nonlinear but smooth vector field or map in the state space implies that the evolution trajectories starting from two close states will still be close to each other in short time, and thus the information of unknown future value can be approximated by the future values of its near neighbors, as sketched in Fig.~\ref{f2}(a). The embedding theory guarantees that such property also holds for states in delay coordinate reconstructed state space. This has inspired the development of making predictions for nonlinear time series using nearest neighbors.

\subsection{Nearest neighbors methods}
The fact that close states will evolve to close future states implies that using near neighbors on the attractor can patch local information to approximate the true future values.
Based on such local patching idea, a large number of creative strategies have been fruitfully proposed to predict future dynamics, including several influential works \cite{Farmer1987predict,Casdagli1989predict,Sugihara1990nature}. The research on this topic has even spawned a competition on nonlinear time series prediction \cite{competition1994}.

To be specific, considering a state $\bm{x}(t)\in\mathbb{R}^d$ on the attractor $\mathcal{A}$ and its $k$ nearest neighbors $\bm{x}(t_i),i=1,2,\dots,k$ where $t_i<t$, the future value $\bm{x}(t+\tau)$ can be locally obtained by the values of $\bm{x}(t_i+\tau)$.   In \cite{Farmer1987predict}, the zeroth-order approximation method was proposed where $\bm{x}(t+\tau)$ was simply approximated by $\bm{x}(t'+\tau)$, i.e.,
\begin{equation}\label{oneneighbor}\bm{x}(t+\tau)\approx\bm{x}(t'+\tau)\end{equation}
where $\bm{x}(t')$ was the nearest neighbor of $\bm{x}(t)$ on the attractor $\mathcal{A}$. In \cite{Casdagli1989predict}, the relation \begin{equation}\label{functionneigbhor}
\bm{x}(t+\tau)=f(\bm{x}(t))
\end{equation} was assumed to be a polynomial and fitted by the $k$-nearest neighbors $\bm{x}(t_i+\tau)=f(\bm{x}(t_i)),i=1,2,\dots,k$.
The simplex method was proposed in \cite{Sugihara1990nature}, where $\bm{x}(t+\tau)$ was approximated by the weighted average of its nearest neighbors' future values in the form of
\begin{equation}\label{weightneighbor}
\bm{x}(t+\tau)=\sum_{i=1}^{k}\omega_i\bm{x}(t_i+\tau),
\end{equation}
where $\omega_i$ was set as an exponential weight to its original distance $\|\bm{x}(t)-\bm{x}(t_i)\|$ and the number of nearest neighbors $k$ was generally set to be larger than the reconstructed dimension $d$ such that $k>d$.
When the future value $\bm{x}(t+\tau)$ is obtained, the prediction of each component $x_i(t+\tau)$ is realized. Particularly, in the case of delay coordinate reconstruction, with the time series data of one dimensional observed variable $x(t)$, the one step prediction of $x(t+\tau)$ is realized, which is the first component of the reconstructed state vector $\bm{x}(t)=[x(t),x(t-\tau),\dots,x(t-(L-1)\tau)]$ in the reconstructed space.

These methods have inspired many data-driven analysis and have been fruitfully reported in the study of multi-disciplines from population study to ecological and biological data even to physiological data and climate data \cite{wu1995balance,kurths1990forecasting,grassberger1991nonlinear,Tsonis1992nature,boyce1992population,longtin1993nonlinear,competition1994,sugihara1994nonlinear,finkenstadt1995forecasting,
ikeguchi1997estimating,hegger1999practical,sello2001solar,smith2002might,judd2003nonlinear,Kantz2004,haefner2005modeling,deyle2013predicting,Ye2016Science,mcgowan2017predicting}.

The research of using local information provided by nearest neighbors to make predictions has been active and attracting even to this day. In order to leverage information from multiple time series observations and overcome the curse of dimensionality,
Ye \textit{et al} proposed a new method in 2016 with name Multiple View Embedding (MVE) \cite{Ye2016Science} which was claimed to be able to make predictions for noisy high-dimensional systems. According to the generalized embedding theorem stated in Theorem \ref{theorem3}, with multiple time series observations composed of $n$ variables and considering $l$ lags for each variable, as many as $m$ possible reconstructed attractors could be constructed by $L-$dimensional variable combinations. Here the number of possible reconstructions could be calculated as  $$m=\left(\begin{matrix}nl\\L\end{matrix}\right)-\left(\begin{matrix}n(l-1)\\L\end{matrix}\right),$$ where $\left(\begin{matrix}nl\\L\end{matrix}\right)$ is the number of possible combinations picking $L$ out of $nl$ coordinates, and it excludes the ``invalid" reconstructions where all $L$ coordinates have positive lags with number $\left(\begin{matrix}n(l-1)\\L\end{matrix}\right)$.

For a specific variable $y$, each reconstruction could be used to make a one-step prediction using nearest neighbors in the way $y_{t+1}\approx y_{nn^i(t)+1}$ where $nn^i(t)$ is the time index of the nearest neighbor in the $i$th reconstructed attractor. Though each single prediction may have reduced precision due to the limited data length and noise, the number of predictions could be fairly large. The MVE method firstly ranks all these predictions according to in-sample forecast accuracy, and then the MVE forecast is defined as a simple average of the top $k$ reconstructions in the way
\begin{equation}\label{MVE}
\hat{y}_{t+1}=\frac{1}{k}\sum^k_{i=1}y_{nn^i(t)+1},
\end{equation}
where $k$ is empirically set to $k=\sqrt{m}$.

The base of information leverage by MVE follows from two aspects. Large number of attractors could be reconstructed due to high-dimensional time series data; and each reconstructed attractor filters the dynamical information of the whole system in a different way. Therefore, combining the information from multiple reconstructed attractors could be advantageous in the manner of Eq.(\ref{MVE}).

\begin{figure}
\centering
\includegraphics[width=0.9\textwidth]{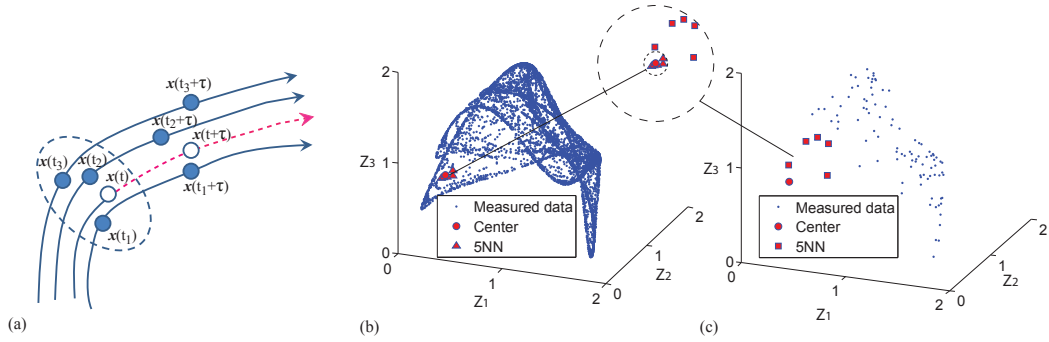}
\caption{(a) Sketch of nearest neighbors and predictions. (b)(c)Nearest neighbors reconstructed by long-term and short-term time series, from Ref. \cite{Ma2014SREP} with permission.} \label{f2}
\end{figure}

Despite of the rapid development of the nearest neighbors based methods in the recent years, one issue on whether nearest neighbors really reflect the local information near the true target is essential to the success of applying such methods. In Fig.~\ref{f2}(b)(c), two attractors are reconstructed for the same system by  different lengths of observed time series. Obviously, when the observed time series is not long enough, the points on the reconstructed attractor are sparse and thus the nearest neighbors actually cannot provide local information of future dynamics. Thus, in order to get the true nearest neighbors, the attractor is necessary to be sufficiently sampled and the time series should be long enough.
  However, real data sets always have a limited length due to the experiments restriction or cost. Therefore, such classic methods are generally not suitable to be used in dealing with short-term data.

On the other hand, the short-term data are usually accompanied with high-dimensional measurements, i.e., the number of state variables $n$ is much larger than the length of measurements $m$. Representative examples of such data are high throughput biological data, e.g. microarray data and gene sequencing data, where tens of thousands of gene probes can be measured simultaneously, but only some tens or even fewer successive time points can be measured for expression of each gene \cite{Miroarray2002} due to various restrictions. Collective dynamics of different complex networks \cite{Strogatz2001Nature} such as social networks and biological networks also show high-dimensional but short time course features, where a large number of coupled dynamical systems evolve via various topological structures but only limited time points can be recorded. As a matter of fact, such kind of high-dimensional biological data have emerged in the big data era, from microscopic gene expression data to mesoscopic neural activity data, even to macroscopic ecological and atmosphere data \cite{HH1952,lockhart2000genomics,de2002modeling,stein2013ecological,fan2014challenges}.

\begin{figure}[t!]
\begin{center}
\includegraphics[width=0.9\textwidth]{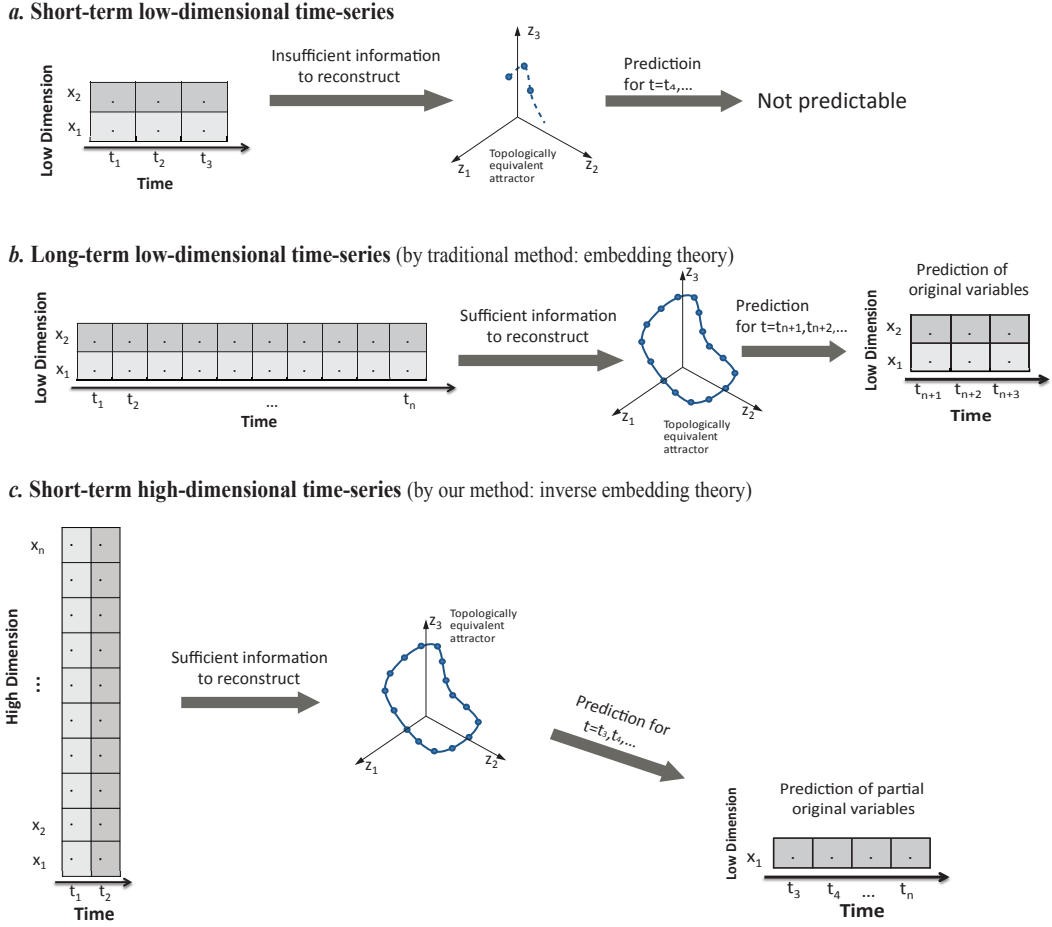}
\caption{Illustrations of information in different types of data sets, from Ref. \cite{Ma2014IJBC} with permission. (a) A short-term
low-dimensional time series has insufficient information to reconstruct the attractor for making predictions. (b) A long-term
low-dimensional time series has sufficient information to reconstruct the attractor based on the embedding theory. Many existing methods are based on such a scheme. (c) A short-term high-dimensional
time series also contain sufficient information, similar to the long low-dimensional time series, which can be used
to reconstruct dynamics or the attractor and thereby make predictions on the time series for some selected variables.} \label{f31}
\end{center}
\end{figure}

In such high-dimensional nonlinear systems, the variables are not independent but intertwined with each other, and the interactions or correlations between them contain rich information contents, which may reflect the accumulated dynamical features of the target system, as illustrated in Fig.~\ref{f31}.  Intuitively, such intertwined relationship, as a source of predictability, has been neglected by the nearest neighbors based methods and
 thus there is a gap between the classic methods and the large amount of big data these days \cite{fan2014challenges}.

\begin{figure}[tb]
\begin{center}
\includegraphics[width=0.5\textwidth]{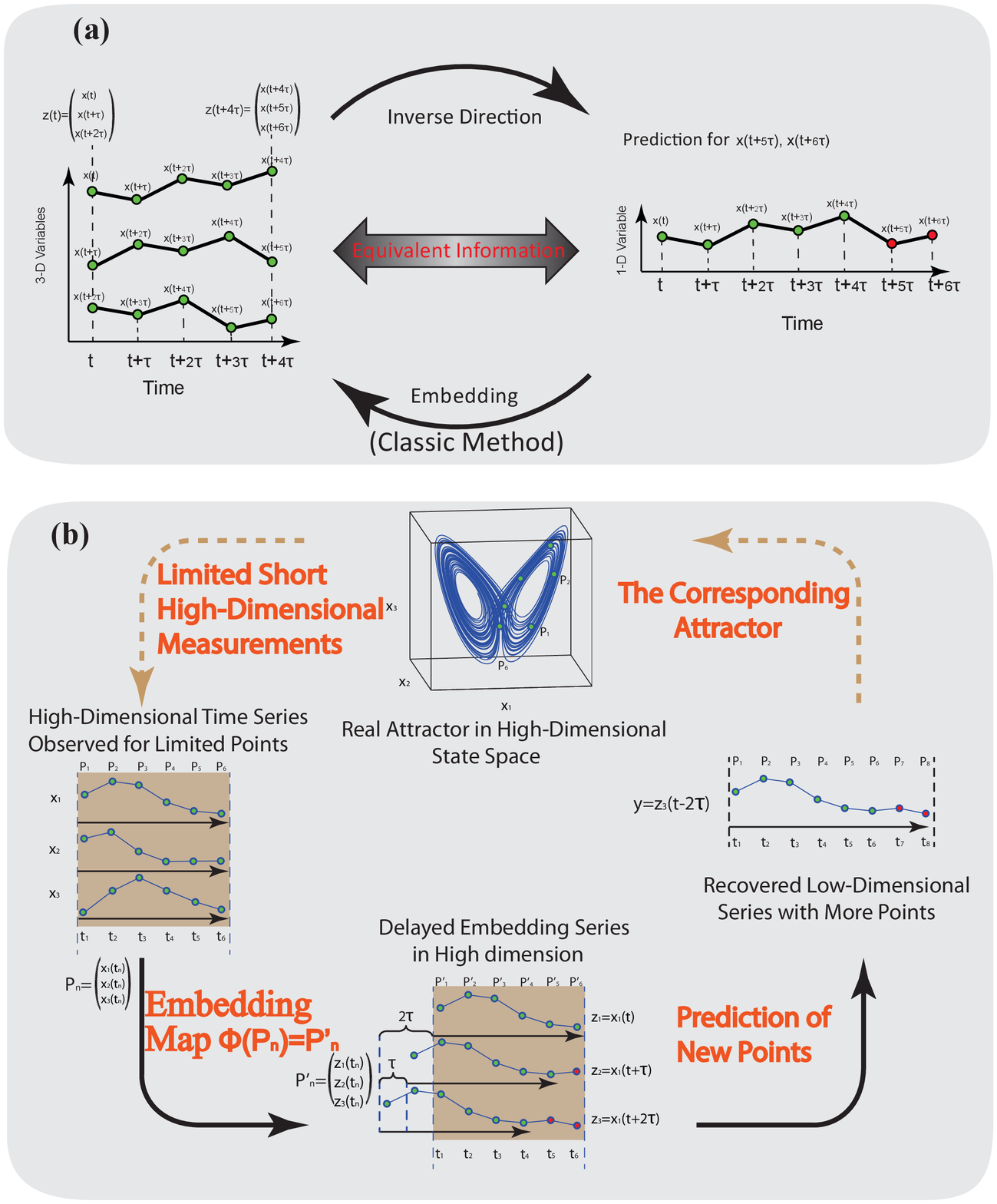}
\caption{Basic ideas of inverse embedding, from Ref. \cite{Ma2014IJBC} with permission. (a) Transformation between short-term high-dimensional time series and long-term scalar
(or low-dimensional) time series.  (b) Sketch of inverse embedding method to predict the dynamics based on short-term
high-dimensional time series data.} \label{f3}
\end{center}
\end{figure}

\subsection{Inverse embedding method}
\begin{figure}[tb]
\begin{center}
\includegraphics[width=0.9\textwidth]{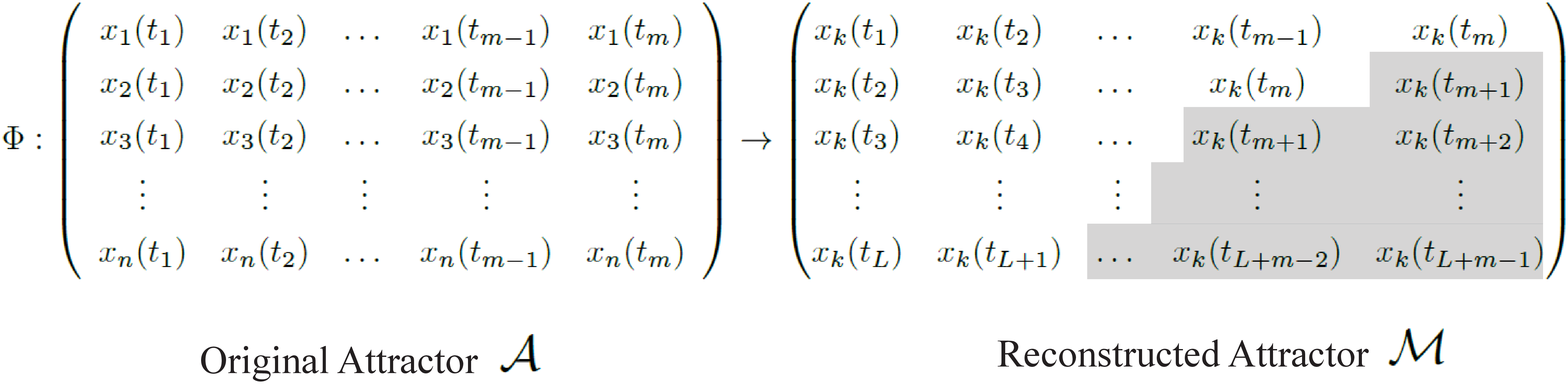}
\caption{Matrix form of the homeomorphism map $\Phi$ between the original high dimensional attractor $\mathcal{A}$ and the reconstructed attractor $\mathcal{M}$, where the shaded variables are values beyond time $t_m$ and play as future predictions. } \label{eq}
\end{center}
\end{figure}

Prediction of short-term high-dimensional data was first addressed in \cite{Ma2014IJBC}, in which Ma \textit{et al} proposed a new SSR based method  aiming to transform the spacial information into a temporal domain and thus made it possible to predict short-term high-dimensional data in a nonlinear manner.
As illustrated in Fig.~\ref{f3}(a), the classic SSR methods start with a single long-term time series and reconstruct a  shorter but higher dimensional time series using delay coordinates. By contrast, the method developed in \cite{Ma2014IJBC} uses the embedding theory in the inverse way, i.e., starting with short high-dimensional time series and recovering the corresponding longer low-dimensional time series.
Theoretically, this new idea can be viewed as an inverse direction of the embedding theory and thus transfers the spacial/interaction information into temporal domain. The process of the inverse embedding method is sketched in Fig.~\ref{f3}(b), and the mapping $\Phi$ between the original high dimensional attractor $\mathcal{A}$ and the reconstructed attractor $\mathcal{M}$ using delayed coordinate, as the key to the method, can be formulated in the form as shown in Fig.~\ref{eq} where the shaded zones stand for the unknown/predicted values.


A simple way to construct the map $\Phi$ is solving it as a linear map, which can be obtained in a very efficient way and can be applied to the prediction of many real systems. However, generally the map $\Phi$ is an unknown nonlinear vector function and the constraints are limited according to Fig.~\ref{eq}. In order to derive the map in an accurate manner so that the transformation can be carried out, it is firstly expanded into a basis functional space, i.e., the power expansion, the RBF network, or the wavelet expansion. Considering the short-term property, the training problem will be generally categorized into an underdetermined problem where there are more coefficients to be determined than the number of independent equations. Note that in many situations, especially with a proper choice of basis functional space, the coefficients in the expanding series can be sparse, thus the recently developed idea of compressive sensing \cite{candes2006stable} can be used to solve this problem.

Using the compressive sensing technique, an algorithm to solve the map $\Phi$ as well as to make predictions was proposed in \cite{Ma2014IJBC}, which was shown to be particularly effective for short-term high-dimensional biological data sets, for instance, microarray data or RNA-seq data, where time points as long as order $O(10)$ were available while thousands of probes were measured and several successful predictions could be made using inverse embedding method. The method has been further considered in the process of big biological data \cite{li2014big} and predicting early-warning signals of critical transitions \cite{liu2015identifying}.

\begin{figure}
\begin{center}
\includegraphics[width=0.7\textwidth]{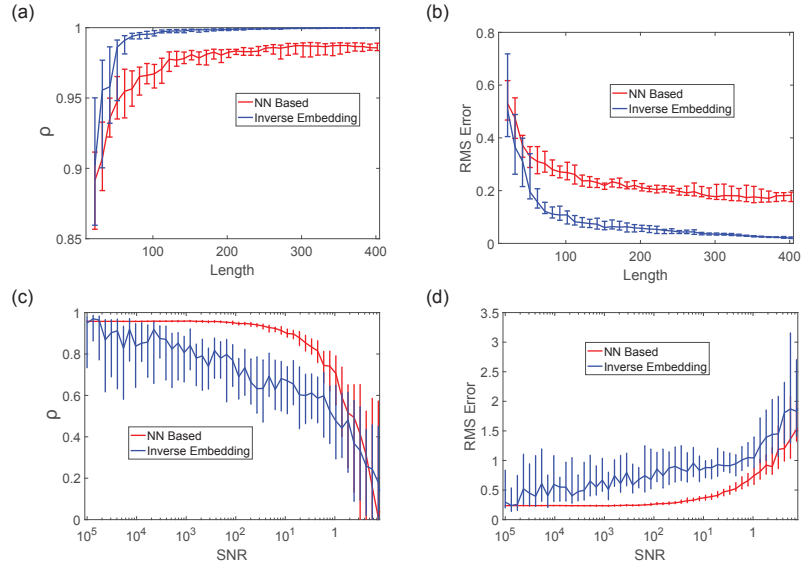}
\caption{Performance comparisons of nearest neighbors (NN) based method and inverse embedding method with different lengths of training data and different levels of noise. Two criteria are used to evaluate the prediction quality: the correlation $\rho$ and the root mean squared (RMS) error between the predicted series and the test data. (a)-(b) Length test is based on $100$ randomly chosen sections for each length of training data. (c)-(d) Noise test is based on $100$ independent trials with time series length $100$. Here the median, the upper quartile, and the lower quartile are shown, respectively. } \label{f4}
\end{center}
\end{figure}

To compare the inverse embedding method and nearest neighbors based method, Fig.~\ref{f4} shows a benchmark test with both methods. A coupled Lorenz system with $15$ variables is considered and two criteria for prediction accuracy are depicted, i.e., the correlation $\rho$ and the rooted mean squared error (RMSE) between the test series and the predicted series. In Fig.~\ref{f4}(a)-(b), it is clearly shown that the inverse embedding method can reach quite a good prediction accuracy with training data of less than $100$ time points, while the nearest neighbor based method is still in the convergence process even with length of $400$. It validates that the inverse embedding method avoids finding nearest neighbors, and thus can deal with short-term data.  In Fig.~\ref{f4}(c)-(d), the robustness of the prediction methods against noise deterioration is also tested. It is shown that the nearest neighbors based method is quite robust against noise, which is mainly attributed to the averaging process of several nearest neighbors. However, compared with the nearest neighbors based method, the inverse embedding method is sensitive to the noise. This is mainly caused by the compressive sensing algorithm adopted in \cite{Ma2014IJBC}, which may enlarge the noise in the process of making parameters sparse. As a matter of fact, though the inverse embedding method transforms the interactions or correlations between variables as spacial information into temporal domain to make predictions, such information actually has redundancy, which is also confirmed in \cite{Ye2016Science}. Therefore, how to dig out such rich redundant information to combat noise could be future work for the inverse embedding method.

\section{Causality detection for nonlinear systems}
\label{sec.Causality}
\subsection{Causal Interactions}
The interactions between variables within one complex system are essential to understand the mechanism behind the experimental phenomenon, and thus are the foundation of mathematical modelling in a system level. Particularly, data driven interaction inference is in an urgent need in order to model various levels of complex networks. The simplest measure of interaction is the correlation between time series data from two variables $x$ and $y$. However, correlation only provides the concomitant covariation between $x$ and $y$ and it has been shown that even a perfect correlation between two observed variables is neither necessary nor sufficient to imply a causal relationship \cite{Sugihara2012Science}, especially when it is recognized that nonlinear dynamics are ubiquitous. Thus, compared to non-directional relationship as correlation, the directional interaction inference is more desired.

Intuitively, the directional interaction from $x$ to $y$ could be understood as the causal effect from the dynamics of $x$ to the dynamics of $y$,  thus the concept of causality can be introduced. Though there still has been no universally accepted definition of causality, various measures of causality or causal relations have been reported and extensively studied. Among them the Granger Causality (GC), proposed by Clive W.J. Granger, the 2003 Nobel laureate in economy \cite{Granger1969Economy,granger2004time},  is undoubtedly a widely accepted definition and it is also a framework to detect causal relationship from time series data. In the sense of Granger Causality, the definition of causality comes from the idea by Wiener in 1956 \cite{wiener1956theory}: ``For two simultaneously measured signals, if we can predict the first signal better by using the past information from the second one than by using the information without it, then we call the second signal causal to the first one". The standard test of GC developed by Granger is based on a linear regression model and later generalized into nonlinear regression \cite{ancona2004radial,marinazzo2008kernel}. Moreover, the original Granger Causality is defined in temporal domain, and the concept could be transformed into frequency domain \cite{brovelli2004beta,kaminski2001evaluating,guo2008uncovering}.
The Granger Causality has been widely accepted in multiple disciplines, especially in economical fields and neural signal analysis, and fruitful applications have been reported in the past decades \cite{geweke1984inference,blinowska2004granger}.
However, one key assumption in Granger Causality is the separation of effect from one variable to another, i.e., the effect of one variable could be separated and removed from the dynamics of the other, which is generally not true in real situations. Particularly, in complex systems when nonlinearity is dominated, every variable could carry all the system's information, as indicated by Takens' embedding theory, and thus the nonlinear systems are generally non-separative \cite{Sugihara2012Science}. Moreover, the Granger Causality is defined for stochastic process, and could not be applied to deterministic systems directly.

\subsection{Nearest neighbors based causality detection}
\begin{figure}[tb]
\begin{center}
\includegraphics[width=0.9\textwidth]{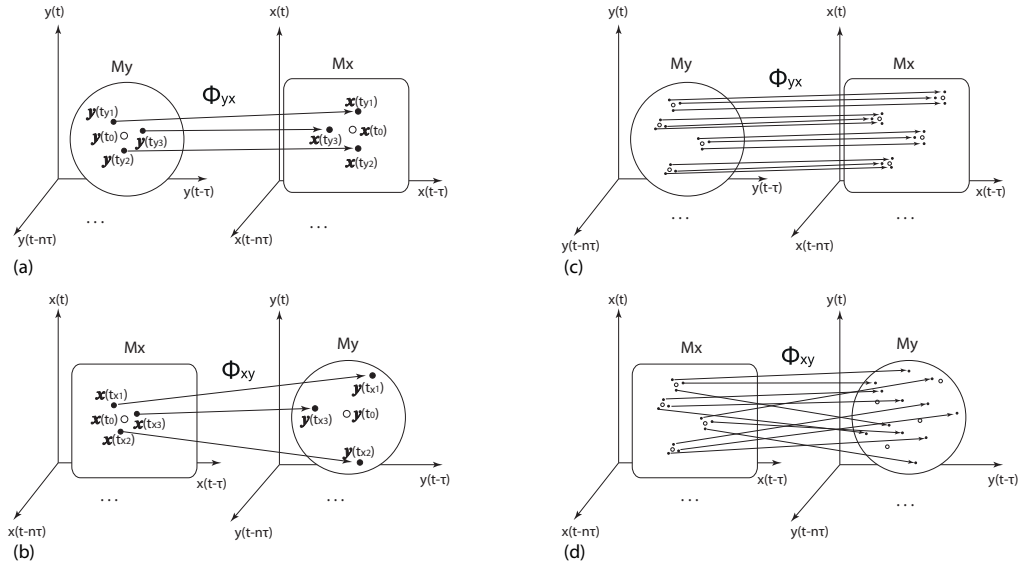}
\caption{Illustrations of mutual neighbors, cross map and smoothness, from Ref. \cite{Ma2014SREP} with permission. (a) For one point $\bm{y}(t_0)\in M_y$ and
 its counterpart $\bm{x}(t_0)\in M_x$, one can find the nearest neighbors $\bm{y}(t_{y_1}),\bm{y}(t_{y_2}),\bm{y}(t_{y_3})$ for
 $\bm{y}(t_0)$ and define the mutual neighbors $\bm{x}(t_{y_1}), \bm{x}(t_{y_2}), \bm{x}(t_{y_3})$ for $\bm{x}(t_0)$. The map between the nearest neighbors and
 mutual neighbors is defined as cross map $\Phi_{yx}$. In the case that $x$ causally influences $y$,
 the cross map $\Phi_{yx}$ maps one neighborhood to one neighborhood. (b) In the case $y$ does not causally influence $x$,
 the cross map $\Phi_{xy}$ does not necessarily map one neighborhood to one neighborhood. (c) and (d) The global smoothness
 of $\Phi_{yx}$ and $\Phi_{xy}$ built from local smoothness. } \label{f5}
\end{center}
\end{figure}

From the viewpoint of information-flow, when $x$ drives $y$, the dynamics information of $x$ is injected into the dynamics of $y$, thus if one can discover such footprint, the driving effect from $x$ to $y$ could be detected. Based on this idea, the state space reconstruction framework provides another choice to consider causality in a purely nonlinear style. Following the embedding theorems, each attractor reconstructed by SSR from delayed coordinates of univariate gives one-to-one relation to the original system's attractor, and thus the two attractors reconstructed by $x$ and $y$ respectively are also one-to-one mapped with each other. Explicitly, such one-to-one map can be set up by the time index, i.e.,
$$\phi:\bm{x}(t)\leftrightarrow\bm{y}(t), \bm{x}(t)\in\mathcal{M}_x,\bm{y}(t)\in\mathcal{M}_y,$$
 where $t$ is the time index as shown in Fig.\ref{f5}(a), and such map $\phi$ is named as a cross map. Moreover, for one particular point $\bm{y}(t_0)\in\mathcal{M}_y$ and the counterpart $\bm{x}(t_0)\in\mathcal{M}_x$, the local neighborhood of $\bm{y}(t_0)$ can be defined on the attractor as $N_\epsilon(\bm{y}(t_0))=\{\bm{y}(t_{y_i})\big|\|\bm{y}(t_{y_i})-\bm{y}(t_0)\|<\epsilon\}$ and the mutual neighbors of $\bm{x}(t_0)$ could be further defined as $$\{\bm{x}(t_{y_i})\big|\bm{y}(t_{y_j})\in N_\epsilon(\bm{y}(t_0))\},$$ as illustrated in Fig.\ref{f5}(a). The mutual neighbors of $\bm{y}(t_0)$ could be analogously defined as show in Fig.\ref{f5}(b).

Intuitively, if $x$ and $y$ are causally linked, the mutual neighbors will have some specific properties due to the information injection. As a matter of fact, in the early development of SSR based causality detection, there are two controversial explanations on how to use such property to define the direction of causality.

In \cite{Quyen1999PhysD}, Quyen \textit{et al} claimed that if $x$ drives $y$ but $y$ has no driving force to $x$, there will be a generalized synchronization relationship between the time series of $x(t)$ and $y(t)$ such that $y=F(x)$ but not vice versa (e.g., no injectivity). Therefore, using the next step of mutual neighbors of $\bm{y}(t_n)\in\mathcal{M}_y$, one can make one-step prediction for $\bm{y}(t_{n+1})$ in the way
$$\hat{\bm{y}}(t_{n+1})=\frac{1}{N}\sum_{j:\bm{x}(t_{x_j})\in N_\epsilon(\bm{x}(t_n))}\bm{y}(t_{x_{j+1}}),$$
where $N$ is the number of nearest neighbors in the $\epsilon-$neighborhood. In such a manner, the cross prediction error ${\rm Pred}_{x\rightarrow y}=\sum\|\hat{\bm{y}}(t_{n+1})-\bm{y}(t_{n+1})\|$ can be used to distinguish the causal relation between $x$ and $y$: if the cross prediction error from $x$ to $y$ is much smaller than the inverse direction, i.e., ${\rm Pred}_{x\rightarrow y}\ll{\rm Pred}_{y\rightarrow x}$, strong evidence of unidirectional causality $X\rightarrow Y$ is implied.

On the other hand, Schiff \textit{et al} proposed another explanation in \cite{schiff1996PRE}. If $x$ and $y$ are unidirectionally coupled, i.e., information flows only from $x$ to $y$, the reconstructed attractor $\mathcal{M}_x$ will present the dynamics in $x$ alone, while the attractor $\mathcal{M}_y$ reconstructed by $y$ will represent the combined dynamics of both $x$ and $y$. Thus, $x$ will be predictable by  cross prediction using mutual neighbors. Obviously, this is completely contradictory in interpreting the differences in cross prediction errors.

This issue has not been resolved until Sugihara \textit{et al} proposed the Convergent Cross Mapping (CCM) method in \cite{Sugihara2012Science}. In \cite{Sugihara2012Science}, the convergence of cross prediction error was firstly stressed as the length of time series data increases, and it was clearly shown that convergence was essential to distinguish the direction of causal link. If the time series data is not long enough, the variance of the cross prediction errors will be quite large and the confidence area of such errors for both directions would overlap with each other, so that it is not possible to distinguish the two directions, as illustrated in Fig.S5 of \cite{Sugihara2012Science}. As the length of time series data increases, the reconstructed manifold becomes denser and the highly resolved attractor improves
the accuracy of prediction based on nearest neighbors, and thus convergence is observed. With the concept of convergence, both theoretical analysis and benchmark test confirm that the direction indicated in \cite{schiff1996PRE} is correct, that is, if $x$ drives $y$, then one can use the information of $y$ to estimate $x$, and vice versa.

Based on the SSR causality detection framework, a lot of applications have been fruitfully reported in multiple disciplines, especially in neurophysiological signal processing, electroencephalographic (EEG) signal processing, brain network reconstruction, ecological data analysis, cardiorespiratory
data analysis, and even climate change study \cite{Arnhold1999PhyD,Quian2002PRE,breakspear2002detection,jamvsek2004nonlinear,stam2005nonlinear,Pereda2005eeg,ansari2006quantitative,causalreview2007,David2008PlosB,nollo2009assessing,ye2015equation,Lindegren13082013,huffaker2014empirically,heskamp2014convergent,egbert2015Nat,mcbride2015sugihara,deyle2016global,tajima2015untangling,muller2016causality,jiang2016directed}.

\subsection{Cross Map Smoothness method}

The convergence concept revealed in \cite{Sugihara2012Science} actually coincides with the argument proposed in \cite{Ma2014SREP,Ma2014IJBC} as shown in Fig.\ref{f2}(b)-(c), that is, using nearest neighbors on the reconstructed attractor as local approximations essentially relies on the density of points on the reconstructed attractor. Only sufficiently long-term time series can bring faithfully reconstructed attractor with dense points on it and consequently the nearest neighbors can be used as local approximations. As shown in Fig. S5 of \cite{Sugihara2012Science}, to realize convergence, a typical length of order $O(10^3)$ was required for a benchmark coupled nonlinear system. Such requirement is usually not practical for typical biological data, e.g.,  high throughput microarray or RNA-seq data for gene expressions of a biological process. Moreover,  on some occasions even though long-term data can be measured, only short (recent) pieces can correctly reflect the causal relation between subsystems due to the nonstationary and fast switching property of the concerned systems. Therefore it is in urgent need of developing new methods to detect causality based on short-term data or a small number of samples.

 \begin{figure}[tb]
\begin{center}
\includegraphics[width=0.9\textwidth]{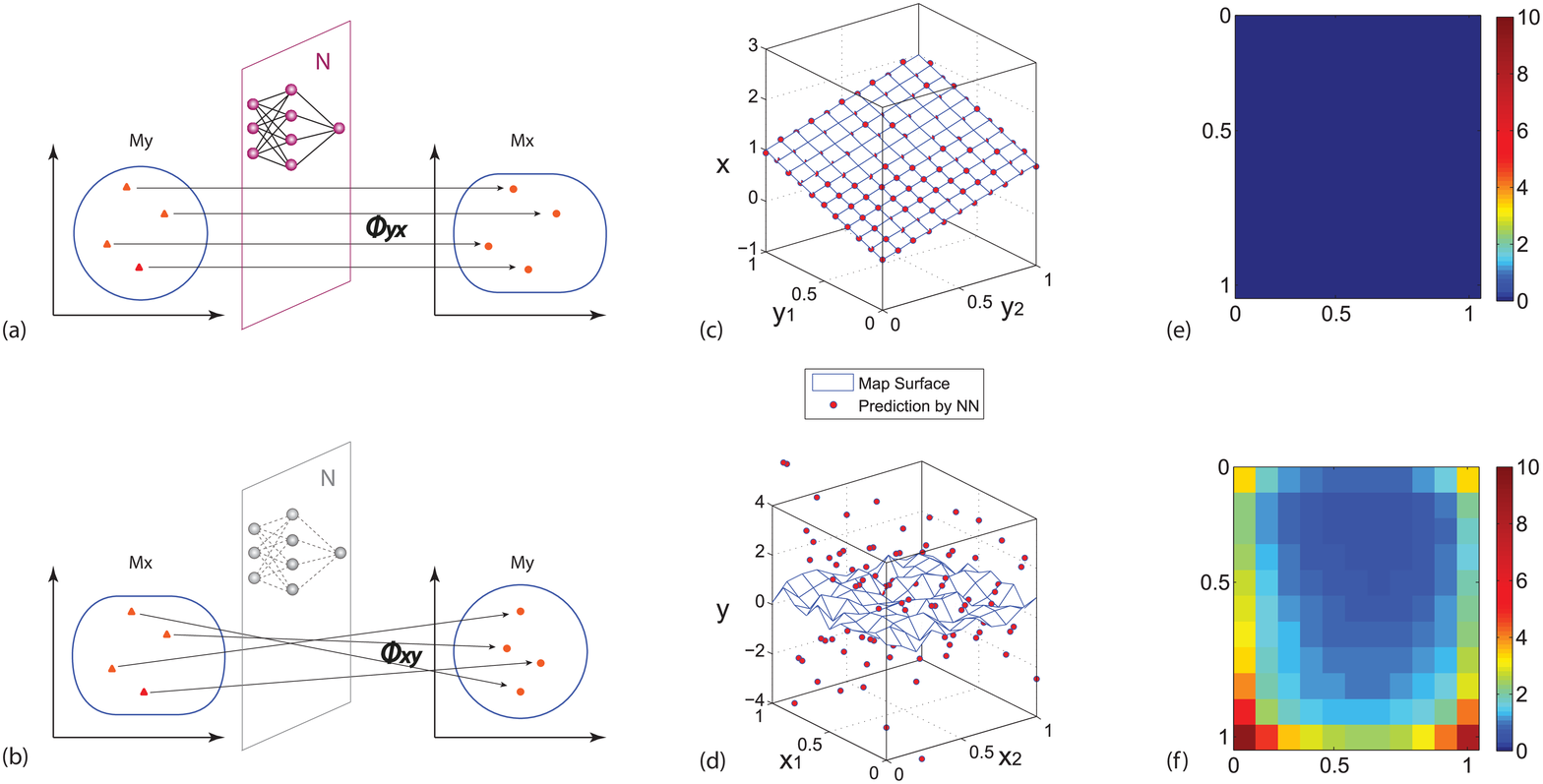}
\caption{Sketch of the cross map smoothness learned by a neural network (NN), from \cite{Ma2014SREP} with permission. (a) and (b) Illustrations
 for the neural network's approximation ability for smooth map and unsmooth map. (c) and (d) The prediction error (or the smoothness of $\Phi$) for cases in (a) and (b) respectively, where
 the leave-one-out scheme is used to calculate errors.
 (e) Assume that $x$ causally influences $y$, the information of $x$ has been encoded in $M_y$
 and consequently $\Phi: M_y\rightarrow M_x$ maps a neighborhood of $y$ to a neighborhood
 of $x$, implying $\Phi_{yx}$ is smooth.  Thus a neural network $\mathcal{N}$ can be trained
 to approximate the map based on the measured data on $M_x$ and $M_y$. (f) Assume that $y$
 has no impact on $x$, then $M_x$ has no information from $y$.  Training a neural network
 to approximate the unsmooth map $\Phi: M_x \rightarrow M_y$ will fail.} \label{f6}
\end{center}
\end{figure}

In \cite{Ma2014SREP}, Ma \textit{et al} proposed a new method named Cross Map Smoothness (CMS) which was also based on the SSR technique but was able to  infer nonlinear causality from short-term time series data. The base of CMS method lies on the observation that the nearest neighbor based methods actually measure the cross map's local property around each point on the reconstructed attractor, leaving the global information unused.  To be specific,  if $x$ causally influences $y$, the mutual neighbors of $\bm{x}(t_0)$ can be used to predict the next value of $\bm{x}$, implying that the nearest neighbors of $\bm{y}(t_0)$ are mapped to close states of $\bm{x}(t_0)$ , i.e., the cross map $\Phi_{yx}: M_y\rightarrow M_x$ is
locally smooth around $\bm{y}(t_0)$, as shown in Fig.\ref{f5}(a). On the reverse direction, if $y$ has no influence over $x$, the image of
$\bm{x}(t_0)$'s neighborhood under the cross map $\Phi_{xy}: M_x \rightarrow
M_y$ is not necessarily the neighborhood of $\bm{y}(t_0)$, thus the cross
map $\Phi_{xy}$ is not necessarily smooth around $\bm{x}(t_0)$, as shown in
Fig.\ref{f5}(b). If the cross map is locally smooth around each point on the attractor, it is further globally
smooth on the whole attractor, and vice versa, as illustrated in Figs.\ref{f5}(c) and (d).
 Thus, the global smoothness
of $\Phi_{yx}$ and $\Phi_{xy}$ can be built from local properties. Moreover, as the coupling
strength increases, information becomes more distinct between the causally
influenced variables. As a result, the reconstructed attractors will contain stronger
historical information from the causes. Thus, finding mutual nearest neighbors is equivalent to measuring the smoothness of the cross map, and the relative smoothness indicates the relative strength of causative effectiveness, i.e., the smoothness of $\Phi_{yx}$
indicates the strength of causative effectiveness from $x$ to $y$.

To quantify such smoothness without finding nearest neighbors, an
efficient algorithm CMS was designed in \cite{Ma2014SREP}. The idea of CMS lies on the fact that any smooth map can be approximated by a neural network $\mathcal{N}$ while training a neural network to approximate an unsmooth map will fail with large training errors, as illustrated in Figs.\ref{f6}(a)-(d). Moreover, the relative training error reflects the relative smoothness, and thus it can be a measure of the relative strength of causative effectiveness. The sketch of the Cross Map Smoothness (CMS) with the neural network (NN) method is illustrated in Figs.\ref{f6}(e) and (f). The test on the benchmark system carried out in \cite{Ma2014SREP} implied that the CMS method could deal with time series data with length around $O(10)$, which greatly reduced the length requirement of $O(10^3)$ by nearest neighbor based methods.

It is stressed that the SSR methods, including both the nearest neighbor based methods and the cross map smoothness method, are designed to detect causality with weak to moderate coupling in nonlinear systems. For too weak coupling, the footprints from the driving force are too low to be measured, while for strong coupling, various kinds of synchronization will appear \cite{mccracken2014convergent}, and causality cannot be defined between synchronized dynamics. Actually, the problem of synchronization, particularly transient synchronization, between two time series will cause problems in causality detection from short-term data, which should be avoided in practical applications.

\subsection{Indirect causality}
\begin{figure}[tb]
\begin{center}
\includegraphics[width=0.7\textwidth]{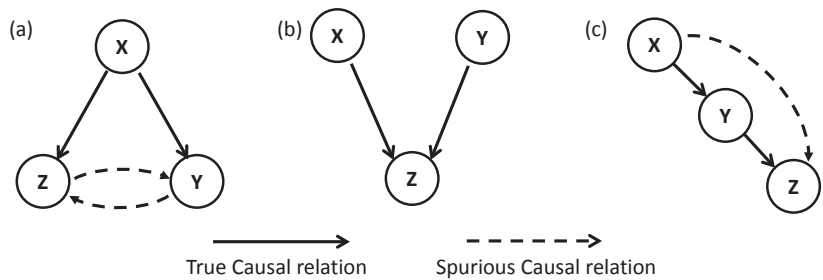}
\caption{Inferring motifs in directional complex network. (a) Fan-out pattern. (b) Fan-in pattern. (c) Cascading pattern.  } \label{f9}
\end{center}
\end{figure}

The causality detection from time series is mainly defined in a pair-wise style, which is the base of network topology inference, but there is still a gap between the existing methods and the network topology reconstruction. Though there are some discussions for the existing methods \cite{Ma2014SREP, Sugihara2012Science}, the detection of complex causality patterns remains uncomplete in the literature.
As indicated in \cite{marbach2010revealing}, three types of motifs are generic in network inference: fan-out pattern, fan-in pattern, and cascading pattern, as shown in Fig.~\ref{f9}.
Intuitively, in the fan-out pattern,  multiple variables are driven by one common source and consequently they are correlated and have the same footprints from the common source, as shown in Fig.~\ref{f9}(a). Both the CCM method \cite{Sugihara2012Science} as well as CMS method \cite{Ma2014SREP} have been shown to work on this pattern, and no significant causality has been detected between the driven nodes. This result confirms the difference between correlation and nonlinear causal relation.
Moreover, Hirata \textit{et al} designed a method specially to deal with common driving source in \cite{hirata2010identifying}  based on the recurrence plot technique, a technique closely related to SSR technique in nonlinear time series analysis. This was demonstrated to work well in climate and brain network reconstruction \cite{hirata2016detecting}. While the fan-in pattern shows no possibility to generate spurious relation, the detected causal strengths are actually weaker as reported in \cite{Ma2014SREP}. The third motif is the cascading pattern in which $x$ drives $y$ and $y$ drives $z$ but $x$ has no direct driving force over $z$, as illustrated in Fig.~\ref{f9}(c). In this pattern, since the information of $x$ has also been injected into the dynamics of $z$ via the intermediate variable $y$, the spurious indirect causality from $x$ to $z$ could be detected by using the SSR methods pair-wisely. Compared to such detected indirect causal relation, it is more desirable to be able to distinguish the direct causal relationship from the indirect causal relationship in the network reconstruction. To the best knowledge of the authors, there is no SSR methods specially designed to deal with this issue to date, which leaves space for future work in the field.

\subsection{Causality with delay}
In the concept of causality, it is generally accepted that ``the cause occurs before the effect" \cite{granger2004time,wiener1956theory}. This implies that the influence of one event on another cannot be instantaneous but delayed after some specific time. As a matter of fact, in the real world, time delays are intimately related to causation, which may result from the transmission time, switching speed, memory effect, or other effects. A real-world example is protein expression regulated by transcription factors, where the regulation process
goes through a cascade of transcriptions and translations that take from minutes to hours to exert their influences. However, the SSR methods, for instance CCM, actually assumes that the time series are on the same footing through the implicit assumption that the directed influence occurs instantaneously, i.e., the detected causality is from $x(t)$ to $y(t)$. Thus, application of the original SSR methods to detect causality may lead to erroneous results. Recently, there are several efforts to incorporate time delay into the CCM scheme.
In \cite{Ye2015SREP}, Ye \textit{et al} considered a single time shift in the underlying time series so as to distinguish time-delayed causal interactions by CCM method. However, it is found that the simple incorporation of time shift into CCM framework is problematic. Specifically, the detected time delay by CCM may drift under different embedding dimensions \cite{Ma2017PRE}, which is definitely confusing. An analogous method was also independently proposed in \cite{schumacher2015statistical}, but it also suffered from the same problem.
Quite recently, this problem has been fixed in \cite{Ma2017PRE}, where a new score Cross Map Evaluation (CME) was designed to substitute the old score in the CCM scheme, and it was reported to be capable of detecting causality with single, multiple, and even distributed time delays. Despite of such progresses in incorporating time delay into SSR based methods, there are still theoretical gaps to be filled. When considering nonlinear systems with time delays, actually one is dealing with infinite-dimensional dynamical systems. However, the classic Takens' embedding theorem is proposed for finite-dimensional systems. Among the few studies on this topic, Robinson has set up an infinite-dimensional Takens theorem in \cite{Robinson1999Nonlin}, which partially gives theoretical foundation for applying SSR method to time delayed systems and sheds light on this problem.

\section{Conclusion}
\label{sec.Conclusion}
The world is ruled by physical laws and each factor is evolving, together with other factors, under various governing laws. Thus the world is complex and no factor could be really independent from each other, but all the factors are intertwined by the dynamics of others.
The state space reconstruction methods, stemming from the embedding theorem, has brought light to the the analysis of such complex and nonlinear world. Particularly, the SSR methods provide a bridge between the observed time series data and the dynamics of the whole hidden original system, making it possible to reconstruct the original system by simply measuring some observed data. Moreover, all these processes could be done in a model-free style and no equations governing the system are required \textit{a priori}. The SSR methods could work as a framework and be effective in various ways of reconstructing the system such as determining dimensionality, characteristic exponent, steady states, unstable periodic orbits, etc. In this review, we focus on two data based applications of the SSR methods: prediction and causality inference for nonlinear systems. Despite the long developing history and fruitful results, these two fields have been still active and attracting even to these days, and new methods keep on getting developed in the recent years. Particularly, in this review, we present in detail the forefront methods to deal with short-term time series data, and several perspectives for future work have been provided. Such developments would create an indispensable tool to big data analysis and systems reconstruction. Also, these types of methods can be applied to the problems of inferring dynamical network structures \cite{zhao2016,liuxp2016} and detecting dynamical criticality of nonlinear systems before the abrupt changes of system states \cite{chen2012,liu2013,liu2015identifying}.

\section*{Acknowledgement}
This research is supported by the National Key Research and Development Program of China (No. 2017YFA0505500), JSPS KAKENHI with grant number JP15H05707, and the National Natural
Science Foundation of China (Nos. 11771010, 31771476, 91530320, 91529303, 91439103, 81471047).


\end{document}